\documentclass[intlimits,twoside,a4paper]{article}
\usepackage[T2A]{fontenc}
\usepackage[cp1251]{inputenc}
\usepackage{graphicx}
\usepackage{amsmath}
\usepackage{amssymb}
\usepackage[eqsecnum]{cmpj}
\issue{2010}{13}{4}{43703}

\title[Analysis of the 3d massive renormalization group
perturbative expansions]%
{Analysis of the 3d massive renormalization group
perturbative expansions: a delicate case
}
\author[B. Delamotte \textsl{et al.}]{B. Delamotte\refaddr{label1}, M. Dudka\refaddr{label2}, Yu. Holovatch\refaddr{label2,label3}, D. Mouhanna\refaddr{label1} }
\addresses{
\addr{label1} LPTMC, CNRS--UMR 7600, Universit\'e Pierre et Marie Curie, 75252 Paris C\'edex 05, France
\addr{label2} Institute for Condensed Matter Physics, National Acad. Sci. of Ukraine, UA--79011 Lviv, Ukraine
\addr{label3} Institut f\"ur Theoretische Physik, Johannes Kepler Universit\"at Linz, A--4040 Linz, Austria
}
\date{Received September 29, 2010}

\begin{document}

\maketitle

\begin{abstract}
The effectiveness of the perturbative renormalization group approach at fixed space dimension $d$ in the theory of critical phenomena is analyzed. Three models are considered: the $O(N)$ model, the cubic model  and the antiferromagnetic model defined on the stacked triangular lattice. We consider all models at fixed $d=3$ and analyze the resummation procedures currently used to compute the critical exponents.  We first show that, for  the $O(N)$ model,  the resummation does not eliminate all non-physical (spurious) fixed points (FPs). Then  the dependence of spurious as well as of the Wilson-Fisher FPs  on the resummation parameters is carefully studied. The critical exponents at the Wilson-Fisher FP show a weak dependence on the resummation parameters. On the contrary, the exponents at the spurious FP as well as its very  existence are strongly dependent on these  parameters. For the cubic model,  a new stable FP is found and its properties depend also strongly on the resummation parameters. It  appears to be  spurious,  as expected. As for  the frustrated models, there are two cases depending on the value of the number of spin components. When $N$ is greater  than a critical value $N_{\mathrm{c}}$, the stable FP shows common characteristic with the Wilson-Fisher FP. On the contrary, for $N<N_{\mathrm{c}}$, the results obtained at the stable FP are similar to those obtained at the spurious FPs of the $O(N)$ and cubic models. We conclude from this analysis that the stable  FP found  for $N<N_{\mathrm{c}}$ in frustrated models is spurious. Since $N_{\mathrm{c}}>3$, we conclude that the transitions for XY and Heisenberg frustrated magnets are  of first order.
\keywords field theory, renormalization group, critical phenomena, perturbation theory, resummation
\pacs 75.10.Hk, 11.10.Hi, 12.38.Cy
\end{abstract}

\section{Introduction}

Renormalization group (RG) methods \cite{wilson74} have  led these last forty years to
such a  deep understanding  of critical phenomena  that it is  hard to
imagine describing them now  without having recourse to these methods.
In  particular, quantitative  computations of  critical  exponents and
universal    amplitude     ratios    have    been     achieved    this
way \cite{field-theory}.  The  prototypical models where  these methods
have  been particularly  successful are  the  three-dimensional ($3d$)
$O(N)$-symmetric systems  modeled by a $\phi^4$  theory that involves
only  one  coupling  constant  \cite{Guida98}.   Such  systems  are  for
instance polymer  chains in a good  solvent ($N=0$)  \cite{deGennes79},
pure ferromagnets with uniaxial anisotropies ($N=1$), superfluidity in
${\rm   He^4}$   ($N=2$),   pure   Heisenberg   ferromagnets   ($N=3$)
 \cite{Pelissetto02},  etc.

The same approach has been used  in the analysis  of $3d$ anisotropic
 \cite{Pelissetto02,Aharony-73}, structurally disordered
 \cite{Pelissetto02,Folk03,Dudka05} and     frustrated     systems
 \cite{Pelissetto02} (see also review in \cite{delamotte04}).   Formally,  these  are described  by
$\phi^4$-like   theories  with   several  coupling constants.   There,  besides
universal exponents  and amplitude ratios, a subject  of interest is
the crossover  between different universality  classes and (universal)
marginal  order  parameter   dimensions  that  govern  this  crossover
 \cite{Dudka04,Holovatch04}.

The prevailing majority of works  in this direction aims at getting an
accurate  description  of  well-established  phenomena.
  As an  example, the  scalar  $\phi^4$ theory  is used  to
obtain  precise  values  of  physical  quantities  characterizing  the
critical behavior  of models belonging to the  $3d$ Ising universality
class. The very  existence of a second order  phase transition in this
model  is a  well established  fact,  observed in  experiments and  MC
simulations and  proven by  analytic tools. On  the other  hand, there
are systems  for which  the very existence  of a second  order phase
transition is  not well established and  RG can be used  to study this
point.  This  is in particular the  case when other  techniques do not
allow  us to conclude.  Technically, if  RG transformations  possess a
fixed point (FP)  which is stable once the  temperature has been tuned
to  $T=T_{\mathrm{c}}$ and  whose basin  of attraction  contains  the microscopic
Hamiltonian, then  the transition  described by this  FP is  of second
order.  However, as  we show in the following, a FP  can be found that
has no physical  meaning and which is only  mathematical artifact.  In
this case, of course, any  conclusion based on this FP is meaningless.
We call it a spurious FP.

As for XY ($N=2$) and  Heisenberg ($N=3$) frustrated magnets, a stable
FP  is found  within  the field-theoretical  perturbative approach  at
fixed  $d$   \cite{Pelissetto01,Pelissetto01a,Pelissetto04,Calabrese02}
which is usually interpreted as  the existence of a second order phase
transition.   However,  this  is  confirmed  neither  by  perturbative
$\varepsilon=d-4$     \cite{kawamura88,antonenko95,calabrese03c}    and
pseudo-$\varepsilon$-expansions  \cite{Holovatch04,calabrese03c} nor by
the    non-perturbative   renormalization   group    (NPRG)   approach
 \cite{tissier00,tissier00b,tissier01,delamotte04}.  The arguments  of \cite{Pelissetto04}, where
such a FP is found,  were questioned recently within minimal subtraction
scheme  \cite{delamotte06,delamotte08}.  In this
paper  we continue the analysis of the non-trivial FPs appearing in the frustrated model along the way
started  in \cite{delamotte06,delamotte08,unpublish}.   Working  within the massive renormalization  scheme we consider
this problem in the general context  of FPs found within the fixed $d$
approach applied  to $\phi^4$-like  theories.  We use  the resummation
method  exploited in  \cite{Pelissetto-cond} and  we study  the
physical  and spurious  FPs  of  the $O(N)$  models  thus showing  the
characteristics of  a spurious  FP. By comparison,  this allows  us to
classify  the FPs  obtained in  the frustrated  models.   A detailed
analysis  of  $O(N)$ and  frustrated  models  was  performed in  $d=2$
elsewhere  \cite{unpublish}. Here, only  the results for the $d=3$ case
are  reported and they  are completed  by results  for the  $3d$ cubic
model.

The paper  is organized  as follows. In  the next section~\ref{II} we
present our  method of analysis  of RG functions  at fixed $d$.  It is
applied to  the $O(N)$  models in $d=3$  in section~\ref{III}, describing
stability of $O(N)$ FPs with  respect to variations of the resummation
parameters and demonstrating differences between physical and spurious
FPs. Then the cubic model is investigated in section~\ref{IIIa}, where
results for the stability exponents at the cubic physical and spurious
FPs are  presented. In section~\ref{IV} frustrated  models with $N=8$,
$N=2$ and $N=3$ are analyzed.  We conclude in section~\ref{VI}.

\section{Perturbative RG at fixed $d$ \label{II}}
Let us start our analysis by introducing the main quantities that
are used within  the field theoretical description of the critical
behavior of the $O(N)$  models. The Hamiltonian  reads:
\begin{equation}
{\mathcal H}= \int{\rm d^d} x \Bigg\{\frac{1}{2}
\left[(\partial\phi)^2 + m_0^2
 \phi^2\right]+
 \frac{u_0}{4!} \left[\phi^2\right]^2 \Bigg \}
\label{landau0}
\end{equation}
 where  $\phi$ is a $N$-component  vector field and
  $u_0>0$ is the  bare coupling.

It is well known that this theory suffers from ultraviolet
divergences. Within the field-theoretical RG approach their removal
is achieved  by an appropriate renormalization procedure followed by a
controlled rearrangement of the perturbative series for RG
functions  \cite{field-theory}. In particular, the change of
couplings under renormalization is obtained from  the
$\beta$-function, calculated as a perturbative series in the
renormalized coupling. The explicit form of this function
depends on the renormalization scheme. However, universal quantities such as critical exponents
 depend neither  on the regularization nor on the renormalization scheme at least if the series expansion
is not truncated at a finite order. A definite scheme must of course be chosen to perform actual calculations.
Among them the dimensional regularization and
the minimal subtraction scheme  \cite{tHooft72} as well as fixed
dimension renormalization at zero external momenta and non-zero
mass (a massive RG scheme)  \cite{Parisi73} are the most used ones.

Introducing a flow parameter $\ell$ (which can be the renormalized mass) the change of the renormalized
coupling constant $u$ under the RG
transformations is by definition given by the equation:
\begin{equation}\label{2}
\ell \frac{\rm d}{{\rm d} \ell} u(\ell) = \beta_u(u(\ell)).
\end{equation}
A fixed point $u^*$ of the  differential equation (\ref{2})
is determined by a  root of the following equation:
\begin{equation}\label{3}
\beta_u(u^*) =  0.
\end{equation}
The FP is said to be stable if it attracts the RG flow when $\ell\to 0$. At the
stable FP
\begin{equation}\label{omeg}
 \omega=\left.\frac{\partial \beta_{u}}{\partial u}\right|_{u=u^*}
\end{equation}
has a positive real part.

The analysis of the perturbative $\beta$-function can
 be performed in two complementary ways. The {\em first
way} consists in obtaining $u^*$, the FP values of the coupling, in the form
of an expansion in a small parameter which is  usually
$\varepsilon=4-d$  \cite{wilson}. In the minimal
subtraction scheme, $\varepsilon$-expansion arises  naturally,
since $\varepsilon$ enters the $\beta$-functions only once as the
 zero-loop contribution. For the  $O(N)$
models and  within the  minimal subtraction scheme the $\beta$-function
is known at
five loops   \cite{kleinert91}. Its two-loop expression is  \cite{note1}:
\begin{equation}\label{6}
\beta_u (u) = -u\bigg(\varepsilon -u+ \frac{3(3N+14)}{(N+8)^2}u^2+\cdots \bigg).
\end{equation}
This function has two roots given by a series expansion in
$\varepsilon$: $u^*=0$ which corresponds to the
Gaussian FP and the non-trivial Wilson-Fisher FP:
\begin{equation}\label{5}
u^*= \varepsilon+{\mathcal O}(\varepsilon^2)
\end{equation}
which is stable for $\varepsilon>0$ and thus governs the critical behavior
of the model for $d<4$.

For the massive scheme, $d$ enters
 all loop integrals and the $\varepsilon$-expansion,
although possible in principle, is in practice performed only at
 low orders. The function  $\beta_u(u)$ is
known in this case at six loops   \cite{antonenko95a} and we give here its two-loop approximation at $d=3$ \cite{note1}:
\begin{equation}\label{mas}
\beta_u (u) = -u\bigg(1 - u + \frac{8(41N+190)}{54(N+8)^2}u^2+\cdots \bigg).
\end{equation}
Instead of the $\varepsilon$-expansion, the
pseudo-$\varepsilon$ expansion is widely used  in the massive scheme  \cite{Nickel}.
It consists in replacing the unity  in the r.h.s. of equation~(\ref{mas})
by the pseudo-$\epsilon$ expansion parameter $\tau$ and in looking for a root of $\beta_u$ as a series expansion in $\tau$. Of course,
$\tau$ must be set equal to one at the end of the calculation.
The non-trivial root  $u^*$ writes:
\begin{equation}\label{7}
u^*= \tau+{\mathcal O}(\tau^2).
\end{equation}
 This technique  avoids a situation when errors coming from the solution of
 an equation for $u^*$ are included into the series for critical exponents.
 Therefore final errors accumulate those coming from
 series for $\beta_u(u)$ and series
 for critical exponents. Instead, the pseudo-$\varepsilon$ expansion results
 in a self-consistent  collection  of  contributions for  the
different steps of calculation  \cite{leguillou80}.  Two common features of the
$\varepsilon$- and  pseudo-$\varepsilon$-expansions are (i)  once
the FP is found  at one loop, it persists at all  loop orders (ii)
in the limit $\varepsilon\to 0$, all FPs coincide with the Gaussian FP
which controls the critical behavior of $\phi^4$ models in $d=4$ (see
 \cite{Kenna93,suslov08} and reference therein).

The {\em  second way} to  analyze the perturbative $\beta$ functions consists in  fixing $d$ and  then in numerically
solving  equations~(\ref{3}).  This   method  is  usually  called  in  the
literature the  fixed-$d$ approach  \cite{Parisi73,schloms87}.  No real
root of  $\beta_u$ can  be {\it a  priori} discarded in  this approach
contrarily to the  $\varepsilon$- and pseudo-$\varepsilon$ expansions,
where  the  only FP  retained  is  by  definition such  that  $u^*\sim
\varepsilon\,\,\mbox{or}\,\,\tau$    \cite{sqrt}.  As  a   result,  the
generic situation is that the number of FPs as well as their stability
vary with  the loop-order: at a  given order, there  can exist several
real and stable FPs or none  instead of a single one. This artifact of
the fixed-$d$ approach is already known and was first reported for the
massive scheme  in $d=3$  \cite{Parisi73}. The  way to deal  with it is
also known: the perturbative  $\beta$-function has to be resummed (see
e.g.   \cite{Hardy}  and  \cite{field-theory}). Resummation  is supposed
both to  restore the Wilson-Fisher  FP when it  does not exist  and to
eliminate the spurious ones. The ability of the resummation procedures
to do so is usually not  questioned.  However, as we show below on the
example  of the  $O(N)$  models,  and then for more general models, spurious FPs  can  exist even  after
resummations have  been performed.  Some extra  criteria are therefore
necessary to eliminate them. We describe these criteria below.

In the minimal subtraction scheme the fixed-$d$ method was introduced
in \cite{schloms87} where
 $\varepsilon$ is kept fixed to 1 and  for the massive scheme calculations
 were performed in $d=2$ and $d=3$ in \cite{Parisi73,Baker76}. We note here, that the
fixed-$d$ approach does not necessarily mean that the space
dimension $d$ is integer: in the minimal subtraction scheme one
can perform calculations for fixed non-integer $\varepsilon$ as
well as in the massive renormalization one can evaluate loop
integrals for non-integer $d$  \cite{Holovatch}.
We explain  in the next subsection our resummation procedure.

\subsection{Resummation}

To extract reliable data from    $\beta$-functions one has to
resum them. Here, following  \cite{leguillou80}, we use the
conformal mapping transform. The main steps are the following. For the
series
\begin{equation}
f(u)=\sum_{n} a_n \ u^n \ \label{series1}
\end{equation}
with factorially growing  coefficients $a_n$ one defines
Borel-Leroy image by:
\begin{equation}
B(u)=\sum_{n} \frac{a_n}{\Gamma[n+b+1]} \ u^n, \ \label{borelsum}
\end{equation}
where $\Gamma[\dots]$ is the Gamma-function. This image is supposed to
converge, in the complex plane, on the disk of radius $1/a$
 where $u=-1/a$ is the singularity of $B(u)$ nearest to the  origin.
Then, using the integral
representation of $\Gamma[n+b+1]$, $f(u)$ is rewritten as:
\begin{equation}
f(u)= \sum_{n} {\frac{a_n}{\Gamma[n+b+1]}}  \ u^n \int_0^{\infty} \
\rd t \,\,\re^{-t}\ t^{n+b} .
\end{equation}
Then, interchanging summation and integration, one can define
the Borel transform of $f$ as:
\begin{equation}
f_B(u)=\int_0^{\infty} \rd t\,\, \re^{-t}\ t^{b}\ B(ut).
\label{boreltrans}
\end{equation}

\begin{figure}[ht]
\begin{center}
\includegraphics[width=12cm]{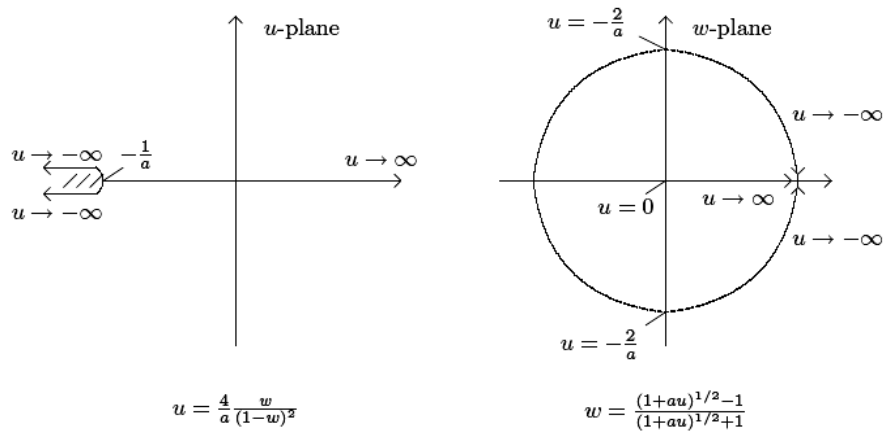}
\end{center}
\caption{ Conformal mapping of the cut-plane onto a disc. See the
text for details. \label{fig57} }
\end{figure}
In order  to compute the integral in  (\ref{boreltrans})  on the
whole real positive  semi-axis one needs  an analytic continuation
of $B(t)$.   It may be achieved by several methods.    In particular,
instead of the series of $B(t)$ its Pad\'e-approximants can be used  \cite{Baker76,baker96,holovatch02}.
We prefer to use here
 a conformal  mapping technique. In this method  assuming that all the
singularities of $B(u)$ lie on the negative real axis and that $B(u)$
is analytic  in the whole complex plane excluding
the cut from $-1/a$ to $-\infty$, one can perform in $B$
the change of variable $w(u)={\sqrt{1 + a\, u}-1\over \sqrt{1 + a\, u}+1}
  \Longleftrightarrow      u(w)={4\over
 a}{w\over(1-w)^2}$, as it is shown in the figure~\ref{fig57}.
This change maps the  complex $u$-plane cut from  $u=-1/a$  to $-\infty$
onto the unit circle in the $w$-plane such that the singularities
of $B(u)$ lying  on the negative axis   now lie on   the boundary
of the  circle $|w|=1$.
Then, the resulting expression of $B(w(u))$ has to be re-expanded in  powers of $w(u)$.
Finally, the resumed expression of  the series $f$ writes:
\begin{equation}
f_{\mathrm{R}}(u)=\sum_{n} d_n(a, b) \int_0^{\infty}
\rd t \,\,{\re^{-t}\, t^{b}\ \left[w(u
t)\right]^n}\ , \label{resummation1}
\end{equation}
where     $d_n(a,b)$   are the coefficients of the re-expansion of
$B(u(w))$ in the powers of $w(u)$.

It is moreover interesting to generalize the above expression (\ref{resummation1}) in the
following way  \cite{kazakov79}
\begin{equation}
f_{\mathrm{R}}(u)=\sum_{n} d_n(\alpha,a, b)  \int_0^{\infty}
\rd t\,\,{\re^{-t}\,  t^{b}}\ { \left[w(u
t)\right]^n \over \left[1-w(u t)\right]^{\alpha} }
\label{resummation2}
\end{equation}
since this allows to impose the strong coupling behavior of the series: $f(u\to \infty)\sim u^{\alpha/2}$.

If  an infinite  number of  terms  for $f(u)$  were known,  expression
(\ref{resummation2}) would  be independent  of the parameters  $a$,
$b$  and  $\alpha$.  However,  once  a  truncation  of the  series  is
performed $f_{\mathrm{R}}(u)$ starts acquiring  a dependence on these parameters.
In principle,  $a$ and $b$  are fixed by  the large order  behavior of the
coefficients $a_n$ in series (\ref{series1})  \cite{leguillou80}, while
$\alpha$ is determined by the  strong coupling behavior of the initial
series. However in many cases, only $a$ is known and $\alpha$ and $b$
must be considered  either as free or variational  parameters.  In any
case, the choice of values of $a$, $\alpha$ and $b$ must be validated a
posteriori by  checking that  a small change  of their values  does not
yield strong variations of the quantities under study.

The above described procedure may be generalized to the case of
several variables. For instance, when $f$ is a function  of two
variables  $u$ and $v$, the  resummation  technique used in  \cite{Pelissetto04}  treats $f$ as a
function of $u$ and of the ratio $z=v/u$:
\begin{equation}
f(u,z)=\sum_{n} a_n(z) \ u^n  \label{series}.
\end{equation}
 Then, keeping $z$ fixed and performing the
resummation only in $u$   according to the steps (\ref{borelsum})--(\ref{resummation2}), one gets:
\begin{equation}
f_{\mathrm{R}}(u,z)=\sum_{n} d_n(\alpha,a(z),b;z)
\int_0^{\infty} \rd t\,\,{\re^{-t}\, t^{b}}{
\left[w(u t;z)\right]^n \over \left[1-w(u
t;z)\right]^{\alpha} } \label{resummation}
\end{equation}
with:
\begin{equation} \label{om}
w(u;z)={\sqrt{1 + a(z)\, u}-1\over \sqrt{1 + a(z)\, u}+1}\
\end{equation}
where,  as     above,  the  coefficients
$d_n(\alpha,a(z),b,z)$ in (\ref{resummation})  are computed  so
that   the  re-expansion of  the right hand side  of (\ref{resummation}) in
powers of $u$ coincides with that of (\ref{series}).

\subsection{Principles of convergence for results obtained with resummation }

The  dependence  of the critical exponents upon the parameters  $a, b$ and $\alpha$ is an indicator
of   the (non-) convergence of the perturbative series. Indeed, in  principle,  any
converged physical quantity $Q$ should be independent of the choice of values of these parameters.  However,
in practice, at a given loop order ($L$), all calculated physical quantities depend (artificially) on them:
$Q\to Q^{(L)}(a,b, \alpha)$. Fixing  $a$  at the value obtained from the large order behavior,
we consider that the optimal result for $Q$ at order $L$  corresponds to the values
$(b_{\rm opt}^{(L)},\alpha_{\rm opt}^{(L)})$ of $(b, \alpha)$  for which Q  depends the least  on $b$
and $\alpha$, that is, for which   it  is stationary:
\begin{equation}
Q^{(L)}_{\rm opt}= Q^{(L)}(b_{\rm opt}^{(L)},\alpha_{\rm opt}^{(L)})\ \ \ \ \ \ {\rm with}\ \ \ \ \ \
\frac{\partial Q^{(L)}(b,\alpha)}{\partial b}\,{\bigg\vert_{b_{\rm opt}^{(L)},\alpha_{\rm opt}^{(L)}}}=\frac{\partial Q^{(L)}(b,\alpha)}{\partial\alpha}\,{\bigg\vert_{b_{\rm opt}^{(L)},\alpha_{\rm opt}^{(L)}}}=0\
\label{PMS}
\end{equation}
where, of course, $b_{\rm opt}^{(L)}$ and $\alpha_{\rm opt}^{(L)}$ are
functions of the order $L$.   The validity of this procedure, known as
the ``Principle of Minimal Sensitivity'' (PMS), requires that there is
a unique  pair $(b_{\rm opt}^{(L)},\alpha_{\rm  opt}^{(L)})$ such that
$Q^{(L)}$  is stationary. This  is generically  not the  case: several
stationary points  are often  found. A second  principle allows  us to
``optimize''  the results  even in  the case  where there  are several
``optimal'' values of  $b$ and $\alpha$ at a given  order $L$: this is
the  so-called ``Principle of  Fastest Apparent  Convergence'' (PFAC).
The idea underlying this principle is that when the numerical value of
$Q^{(L)}$ is  almost converged (that  is $L$ is sufficiently  large to
achieve a  prescribed accuracy) then  the next order  of approximation
must consist  only in a  small change of this  value: $Q^{(L+1)}\simeq
Q^{(L)}$. Thus, the preferred values of $b$ and $\alpha$ should be the
ones  for   which  the   difference  between  two   successive  orders
$Q^{(L+1)}(b^{(L+1)},\alpha^{(L+1)})-    Q^{(L)}(b^{(L)},\alpha^{(L)})$
is minimal.  In  practice, the two principles should  be used together
for consistency and, if  there are several solutions to equation~(\ref{PMS})
at  order $L$  and/or $L+1$,  one should  choose the  couples $(b_{\rm
opt}^{(L)},\alpha_{\rm         opt}^{(L)})$        and        $(b_{\rm
opt}^{(L+1)},\alpha_{\rm  opt}^{(L+1)})$   for  which  the  stationary
values   $Q^{(L)}(b_{\rm   opt}^{(L)},\alpha_{\rm   opt}^{(L)})$   and
$Q^{(L+1)}(b_{\rm  opt}^{(L+1)},\alpha_{\rm   opt}^{(L+1)})$  are  the
closest,  that is  for which  there is  fastest  apparent convergence.
These    principles    have     been    developed    and    used    in
 \cite{leguillou80,mudrov98c,delamotte08}, { see also  \cite{Dudka04}}.

\section{$O(N)$-symmetric theory \label{III}}
\begin{figure}[ht]
\centerline{\includegraphics[width=0.4\textwidth]{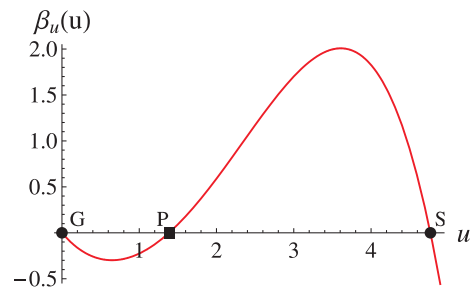}}
\caption{Resummed  $\beta_u(u)$-function of the $O(4)$ model
at $\alpha=6$ and $b=4$. The disks denote unstable  FPs {\bf G} and {\bf S}, while
the square corresponds to the  stable FP {\bf P}.} \label{O(m)}
\end{figure}

Taking the above resummation procedure, let us show that it does not always eliminate
unphysical FPs even in the simplest $\phi^4$ model in $d=3$.  We study the $\beta_u$-function
obtained within the massive scheme at four-, five- and six-loop orders  \cite{antonenko95a}.
As was noticed in  \cite{Parisi73},  the non-resummed $\beta_u$-function has no
non-trivial root at the even orders of perturbation theory.
Applying the resummation procedure described by equations~(\ref{series1})--(\ref{resummation2}) we find
 the Wilson-Fisher FP {\bf P} close to the origin (see figure~\ref{O(m)}). There
 exists a large amount of studies of  FP {\bf P} and its properties are
 well-known (see e. g. \cite{field-theory,Guida98,Pelissetto02}  and references therein).
 However, we also find for larger  $u$,  in addition to {\bf P} and for  some values of  $\alpha$
and $b$, a new FP  that we call {\bf S}  (in even orders of the loop
expansion), see
figure~\ref{O(m)}. Although this FP is unstable  it changes the RG
flow structure and, if taken seriously, it would correspond to a tricritical FP (\ref{landau0}). However for the
model under consideration the FP structure is well established
 \cite{field-theory}  and this additional FP must be considered as an artifact of the
 fixed-$d$ approach.

Let us compare the stability properties
of the critical exponents defined at  {\bf P} and {\bf S}. Here,  we consider
  $\omega$ defined in equation~(\ref{omeg}), which is
the leading correction to scaling exponent.
We compute it at  {\bf P} and {\bf S} from the  resummed $\beta_u$-function
and study its stability with the loop-order $L$  and with respect to variations of the resummation parameters  $b$ and
$\alpha$.
Note that although the large order behavior of the series is known and leads to a preferred value
of $a$   \cite{Lipatov77,Brezin77} it is  possible to take it as a free parameter and to also study
the stability of all the results w.r.t. variations of this parameter. We choose here to fix $a$, then to vary $b$ and $\alpha$
so as to satisfy the PMS and PFAC and finally to check  the stability of our results with respect to small variations of $a$.

\subsection{ Wilson-Fisher FP }

First we consider $\omega$ for
the Wilson-Fisher FP {\bf P}. For $O(N)$ models, the   analytically
calculated value of $a$ is known to be $a=0.14777422\times
9/(N+8)$  \cite{leguillou80} and we  use it in our analysis.

\begin{figure}[ht]
\hspace{5mm}
\includegraphics[width=0.45\textwidth]{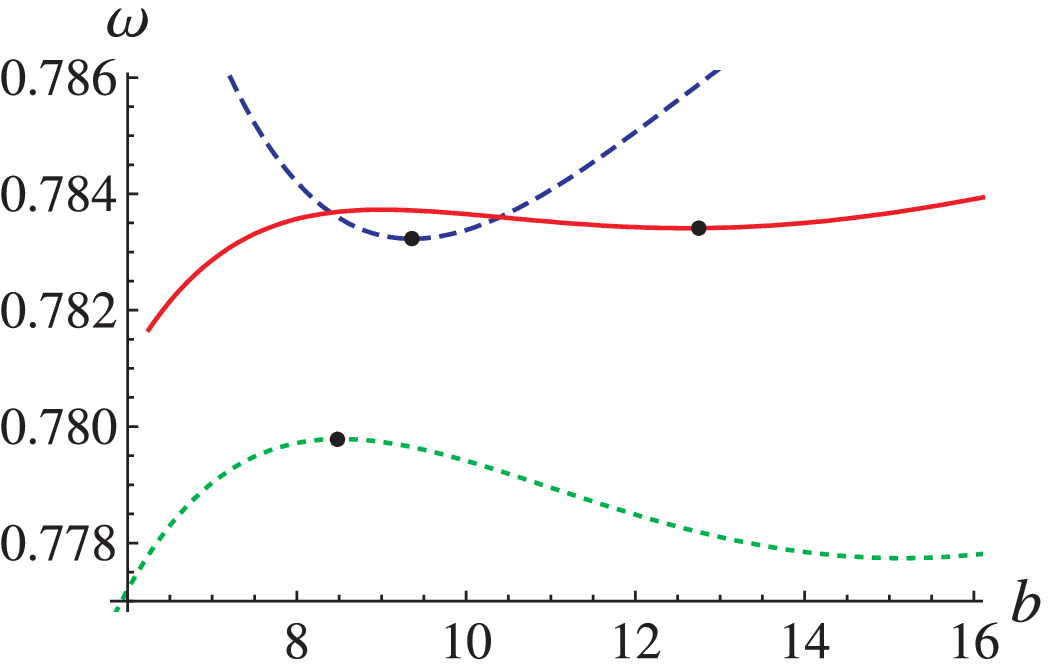}%
\hfill%
\includegraphics[width=0.45\textwidth]{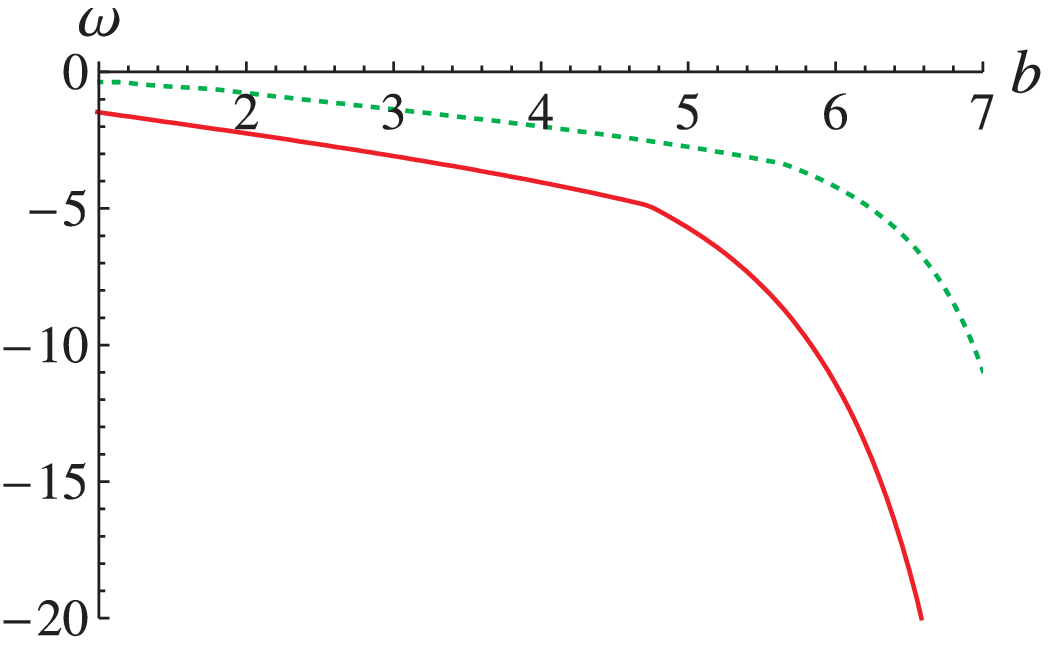}
\hspace{5mm}%
\\%
\parbox[t]{0.48\textwidth}{%
\centerline{(a)}%
}%
\hfill%
\parbox[t]{0.48\textwidth}{%
\centerline{(b)}%
}%
\caption{The exponent $\omega$ of the three-dimensional $O(4)$ model as a function of the
resummation parameter $b$
calculated at (a) FP {\bf P}  and (b) FP {\bf S}. Results for  FP {\bf P} are presented
at  four- (grey dashed curve (green online)), five- (dark dashed curve (blue
online)) and six-loop (solid curve) orders. The dots
at each loop-order correspond  to a stationary value of
$\omega=\omega(\alpha,b)$ in both  $\alpha$ and $b$ directions at the same time ($L=4$:
$\alpha=3.2$, $b=8.5$, $L=5$: $\alpha=5.2$, $b=9.5$, $L=6$: $\alpha=5$, $b=13$), at these values  fastest apparent convergence between two
successive loop orders is observed. Results for the FP {\bf S} are presented in six and four loops at $\alpha=6$. No stationary point is found. The exponent $\omega$ is negative since {\bf S}
 is repulsive.}
\label{omega_O}
\end{figure}
The results of our calculations are shown in  figure~\ref{omega_O}~(a)  where the exponent $\omega$  for $N=4$ ($a=0.110831$) is shown  as a function of the parameter $b$  for the values of $\alpha$  for which  stationarity  is found for both $b$ and $\alpha$. The values obtained for four-, five- and six-loop orders are very  close. An indicator of the quality of the convergence is given by the difference between the fifth and the sixth orders:  $\omega(L=6)-\omega(L=5)\simeq  2.  10^{-4}$.  Our six loop estimation is  $\omega\simeq  0.783$. This result is  compatible with the  results obtained  by Guida and Zinn-Justin  \cite{Guida98} for $N=4$: $\omega=0.774\pm0.020$ (fixed $d=3$, six/seven loops),  $\omega=0.795\pm0.030$ ($\epsilon$-expansion, five loops), which is a check of the reliability of both the resummation scheme and the convergence criteria used here.

\subsection{Spurious FP}
We now analyze $\omega$ at the  FP $\bf S$. As expected, the results  computed at  $\bf S$ are very
unstable with respect to variations of the parameters $\alpha$ and $b$.
No stationary point is found for both $b$ and $\alpha$. To show the dependence  of $\omega$ on $b$ at
different loop orders we thus choose a given typical value of $\alpha$, see figure~\ref{omega_O}~(b). A monotonous
decrease of $\omega$ while increasing  $b$ is found.  At fixed values
of $\alpha$ and $b$ we moreover find that  $\vert\omega\vert$  increases with the loop order.

At the end of this section some conclusion  from the fixed
$d=3$ analysis of the $O(N)$ model can be made. Our results for
$\omega$ at  {\bf P} and  {\bf S} demonstrate two different
features. For {\bf P},
the convergence principles that we use --- PMS and PFAC --- allow us to find a reliable value
of $\omega$, which is consistent with other estimations.
On the contrary,  $\omega$ at {\bf S} does not show any  stationarity when $b$ and $\alpha$ are varied.
Furthermore, large differences between values in different loop orders (comparing to the results for FP {\bf P}) are observed.

\section{Cubic anisotropy \label{IIIa}}

The previous section was devoted to the study of the $\phi^4$ model with one
coupling.
We now consider a more complicated model with two couplings,
namely describing cubic  anisotropy.
The effective Hamiltonian for this model reads
 \cite{Aharony-73}:
\begin{equation}
{\cal H} = \int \rd^d x
\left\{ {1\over 2}
      \left[ (\partial \phi)^2 +  m_0^2 \phi^2 \right] +\frac{u_0}{4!} \left[\phi^2\right]^2
  + \frac{v_0}{4!} \sum_{i=1}^N \phi^4_i  \right\}.
\label{cubic_mod}
\end{equation}

The Hamiltonian
(\ref{cubic_mod})  is  used  to  study the critical   behavior of numerous
magnetic  and  ferroelectric systems  with appropriate order parameter
symmetry (see e.g.   \cite{folk00b}).   The $\beta$-functions are known
  at six-loop    order  in the   massive  scheme
 \cite{carmona00}. Their FPs structure is known  \cite{Aharony-73}. The   FP $\bf G$ ($u^*=v^*=0$) and the  Ising
FP ($u^*=0,  v^*\ne 0)$ are  both  unstable for all values  of $N$. Two other  FPs: the $O(N)$ symmetric FP {\bf P}
$(u^*\neq0,  v^*=0)$ and the mixed one   $\bf M$ $(u^*\neq 0, v^*\neq 0)$ interchange their stability
at a critical value $N_{\mathrm{c}}$ of  $N$: for $N<N_{\mathrm{c}}$ the FP {\bf P} is stable and
$\bf M$ is unstable and vice versa  for $N>{N_{\mathrm{c}}}$.  The  critical
value ${N_{\mathrm{c}}}$ has been found  to  be slightly  less than 3:  for
instance ${N_{\mathrm{c}}}= 2.89\pm 0.04$ in
 \cite{carmona00} and ${N_{\mathrm{c}}}= 2.862\pm 0.005$ in  \cite{folk00b}.

\begin{figure}[ht]
\hspace{5mm}
\includegraphics[width=0.45\textwidth]{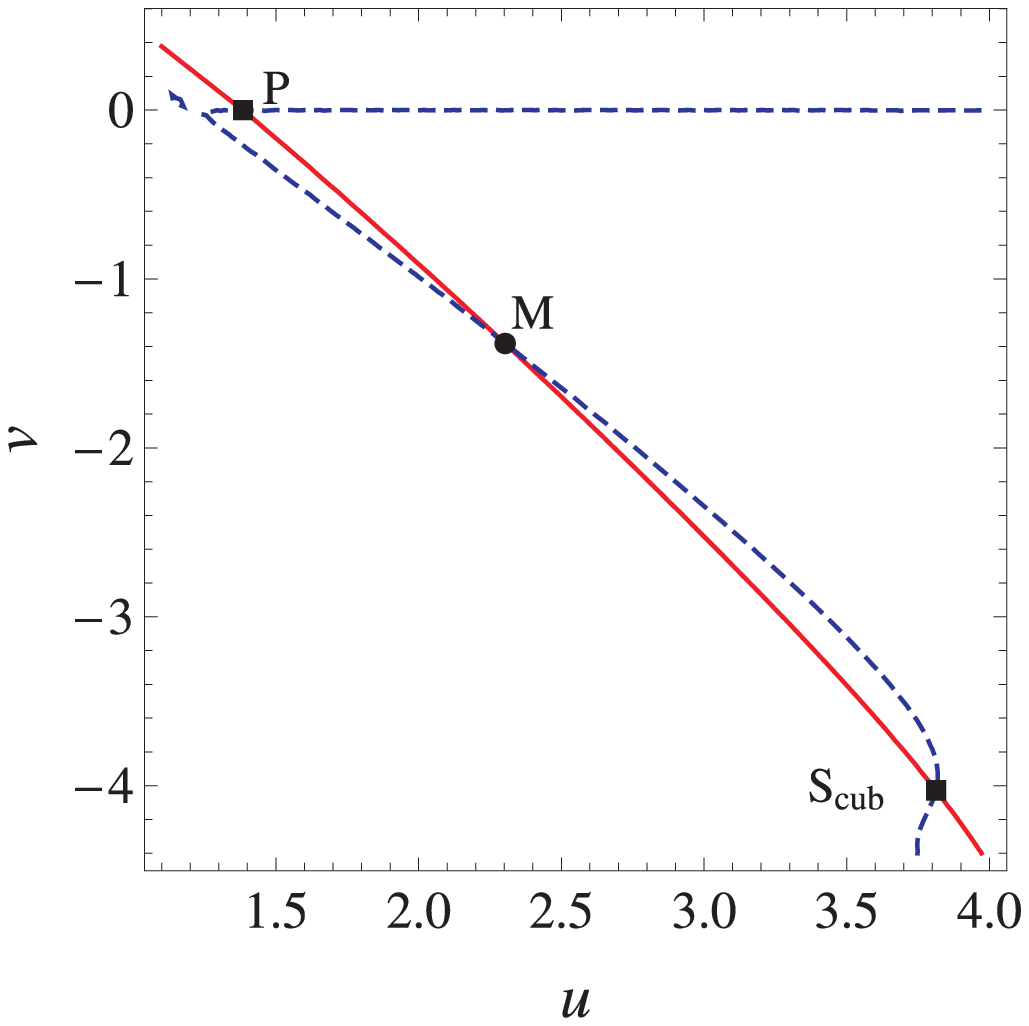}%
\hfill%
\includegraphics[width=0.45\textwidth]{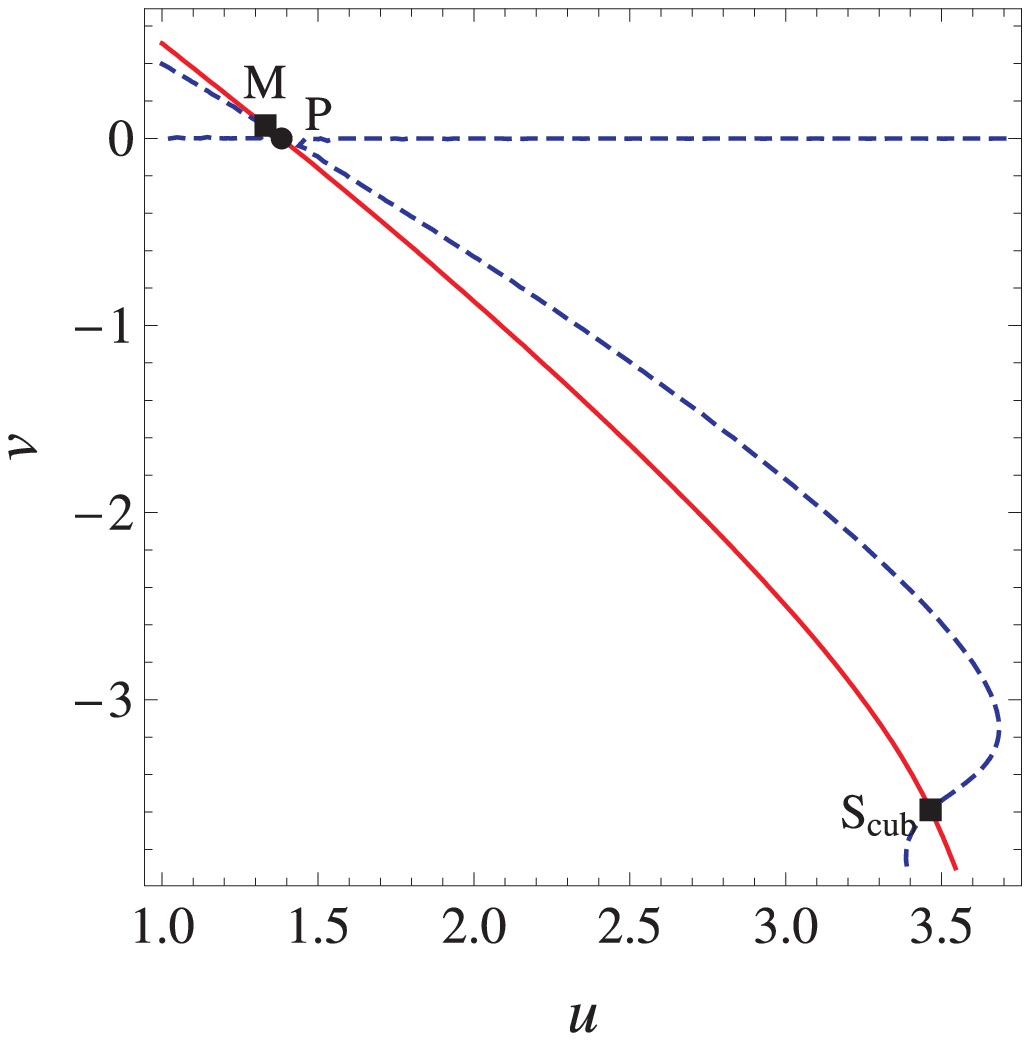}
\\%
\parbox[t]{0.48\textwidth}{%
\centerline{(a)}%
}%
\hfill%
\parbox[t]{0.48\textwidth}{%
\centerline{(b)}%
}%
\caption{Lines of zeroes of $\beta_u$ (solid curves) and
$\beta_v$ (dashed curves) for the cubic model for (a) $N=2$ with $\alpha=9$ and $b=9$ and (b) $N=3$ for $\alpha=8$  and $b=10$.
The intersections of the  solid and dashed lines
correspond to different FPs:  Wilson-Fisher
{\bf P}, cubic {\bf M} and spurious
{\bf S}$_{\rm cub}$. The discs correspond to unstable FPs, while squares denote stable FPs.}
\label{cubic}
\end{figure}
We have performed a stability analysis  for the $N=2$, $N=3$ cases analogous
to the $O(N)$ case. We
take the $3d$ six-loop massive functions  \cite{carmona00}
and apply resummation procedure generalized to the case of two
variables (\ref{series})--(\ref{om}). Now, the FPs coordinates  are defined by the system of equations:
\begin{equation}
 \left\{
  \begin{array}{l}
   \beta_u(u,v)=0,\\
   \\
   \beta_v(u,v)=0.
  \end{array}
 \right.
\end{equation}
Here, $ \beta_u$ and $\beta_v$ stand for the resummed
$\beta$-functions of the model defined by~(\ref{cubic_mod}).
Being functions of the two variables $u$ and $v$,
$\beta_u$ and $\beta_v$ describe surfaces in the space $(u,v,\beta)$. The FPs
correspond to the points of common
intersections of these two surfaces with the $(u,v)$-plane. As a guide for the eyes, we  plot the lines of zeroes of the resummed
 $\beta_u$ and $\beta_v$ functions. The FPs, if they exist,
correspond to the crossing points of such lines. To obtain such
curves, one must use definite values of the resummation
parameters $a,\,b,\,\alpha$. As it has been shown in
 \cite{Pelissetto01} the  parameter $a$ for the cubic model (\ref{cubic_mod})
depends on the ratio $z=v/u$ and is:
$a=0.14777422\times (9/(2N+8)+z)$ for $z>0$ and
$a=0.14777422\times( 9/(2N+8)+z/N)$ for $-{2 N\over (N+1)} \times{9\over (N+8)} <z<0$.
The resummed $\beta$-functions with the value $a$
 above are depicted in figure~\ref{cubic} for $N=2$ and $N=3$.

From  figure~\ref{cubic} one
  observes that, in addition to the  usual
FPs {\bf P} and {\bf M}, there exists another stable FP  that   has     no    counterpart   within     the
$\varepsilon$-expansion for both values of $N$, denoted by us as {\bf S}$\rm _{cub}$.
The   presence of this FP,  if   taken seriously, would have important
physical  consequences  since it would correspond   to  a second order
phase transition    with a new universality   class.   However no such
transition has  ever been reported.   On the  contrary, a first  order
behavior for  all values of   $N$ larger than  ${N_{\mathrm{c}}}$ and $v_0<0$ is found
within perturbative
 \cite{carmona00,calabrese03d} or non-perturbative  \cite{tissier01b}
field theoretical analysis as well as numerical simulations
 \cite{itakura99} in related systems (four-state antiferromagnetic
Potts model). Below we consider the dependence of physical
quantities for the FP {\bf M} and the spurious FP {\bf S}$\rm _{cub}$ upon
variations  of the resummation parameters. We then compare the results obtained with those for the $O(N)$ model.

\subsection{Mixed cubic FP}
Since the cubic model involves two coupling constants $u$ and $v$
the stability of its FPs is defined by two  exponents,
$\omega_1$ and $\omega_2$, which are the generalizations of equation~(\ref{omeg}) for the case of two variables.
\begin{figure}[htbp]
\hspace{5mm}
\includegraphics[width=0.45\textwidth]{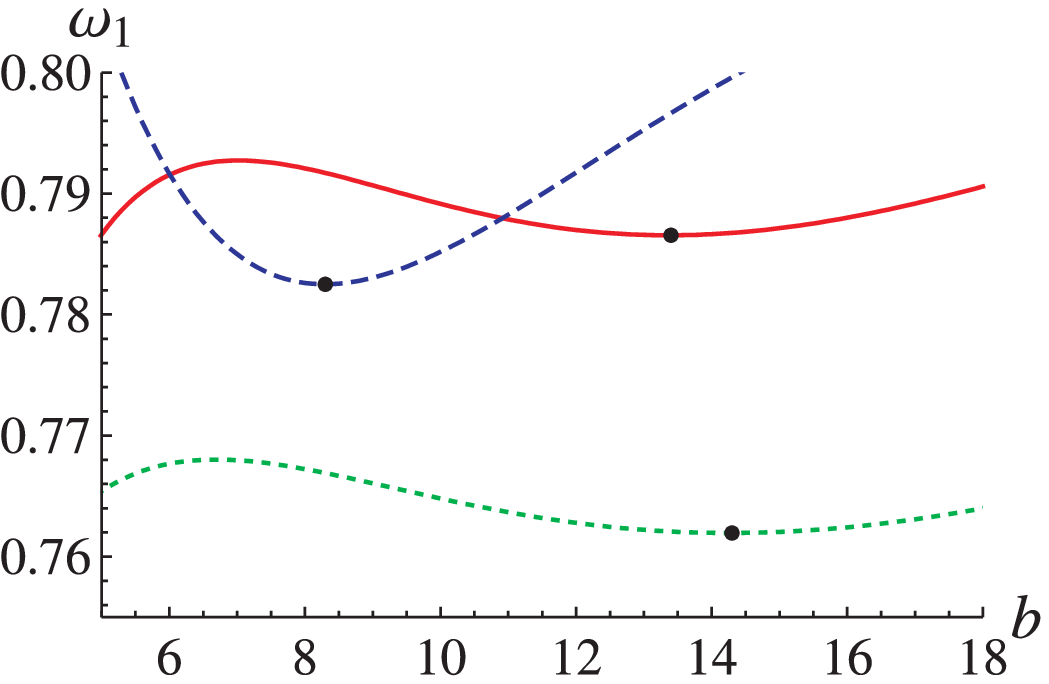}%
\hfill%
\includegraphics[width=0.45\textwidth]{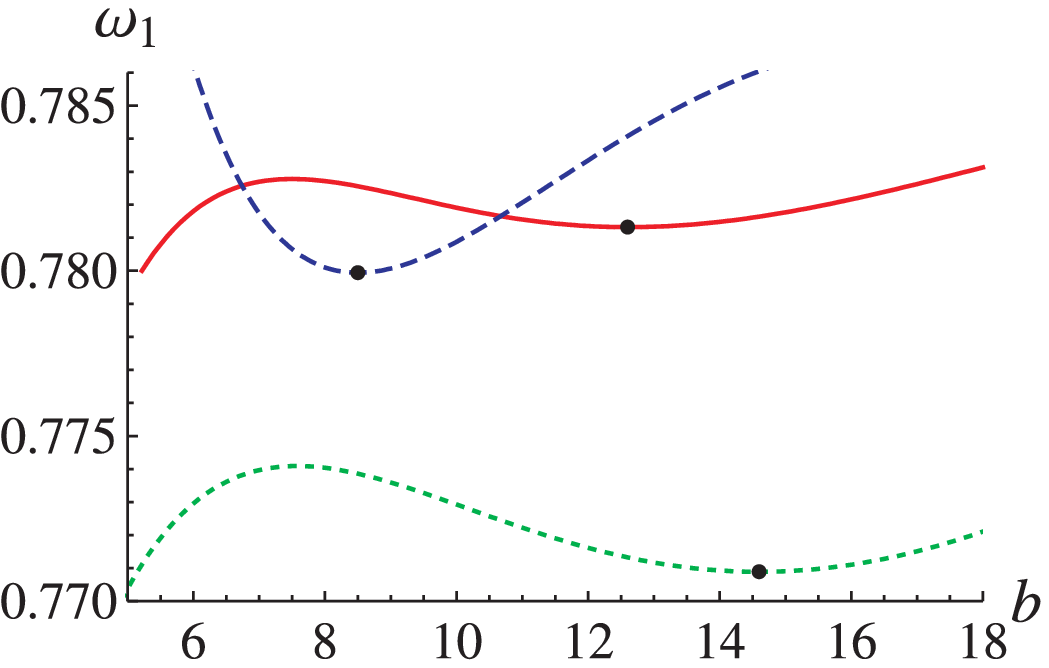}
\hspace{5mm}%
\\%
\parbox[t]{0.48\textwidth}{%
\centerline{(a)}%
}%
\hfill%
\parbox[t]{0.48\textwidth}{%
\centerline{(b)}%
}%
\caption{The exponent $\omega_1$ at the mixed FP {\bf M} for  $N=2$ (a)
and  $N=3$ (b) as a function of  $b$   at four
(grey dashed curves (green online)), five (dark dashed curves (blue
online)) and six (solid curves) loops.
The dots at each loop order correspond  to a stationary value of
$\omega=\omega(\alpha,b)$ in both  $\alpha$ and $b$ directions at the same time. For $N=2$:
$\alpha=3.4$, $b=14.3$ ($L=4$),  $\alpha=5.7$, $b=8.3$ ($L=5$),  $\alpha=5.8$, $b=13.4$ ($L=6$); for $N=3$: $\alpha=3.3$, $b=14.6$ ($L=4$),  $\alpha=5.3$, $b=8.5$ ($L=5$),  $\alpha=5.2$, $b=12.6$ ($L=6$). At these values  fastest apparent convergence between two
successive loop orders is observed.}
\label{cub3}
\end{figure}
At a  stable FP,  both $\omega_1$ and $\omega_2$ have  positive real parts.
Calculating these exponents for FP {\bf M} (varying $\alpha$ and $b$ in different $L$) we observe a similar picture
as the one obtained for $\omega$ computed at {\bf P} in $O(N)$  models (which is intuitively expected).
In  figure~\ref{cub3} we only present the  behavior of the larger exponent $\omega_1$ as
 function of the resummation parameter.
Stationary points in $\alpha$ and $b$ are found. The values calculated  at different loop orders
at these points  are very close to each other. For instance, for $N=3$ $\omega(L=6)-\omega(L=5)\approx  0.001$. Our six-loop results
  $\omega_1=0.781$, $\omega_2=0.011$ for $N=3$ agree with other six-loop estimates $\omega_1=0.781\pm0.004$, $\omega_2=0.010\pm0.004$  \cite{carmona00} and $\omega_1=0.777\pm0.009$, $\omega_2=0.015\pm0.002$  \cite{folk00b,note_cub}.

\subsection{Spurious cubic FP}
\begin{figure}[htbp]
\hspace{5mm}
\includegraphics[width=0.45\textwidth]{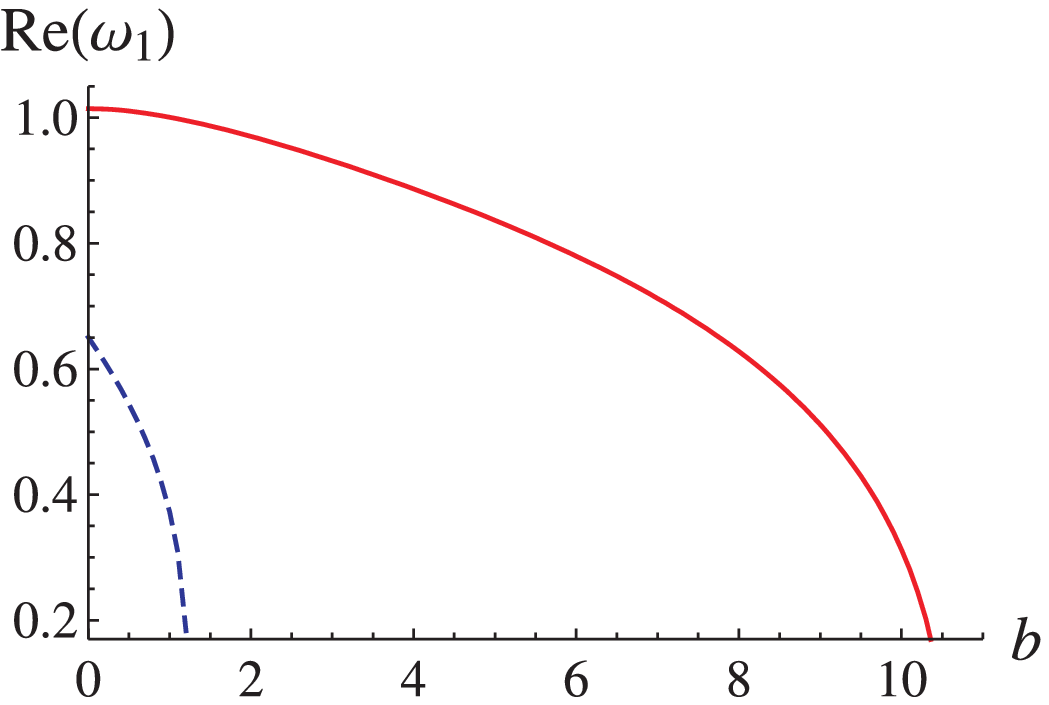}%
\hfill%
\includegraphics[width=0.45\textwidth]{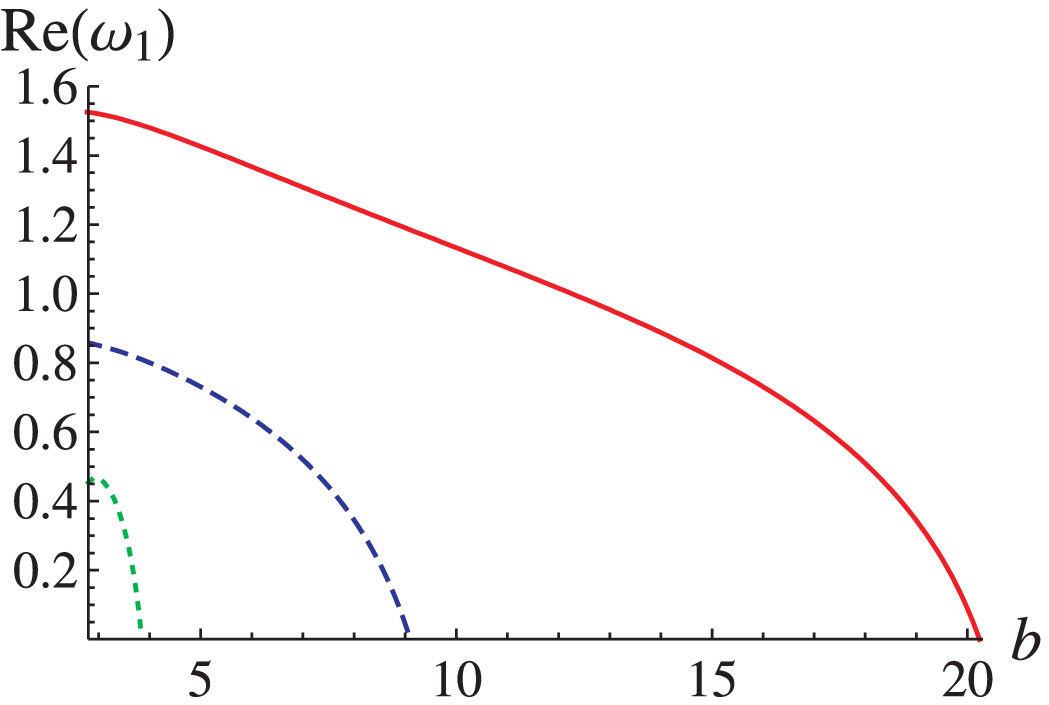}
\hspace{5mm}%
\\%
\parbox[t]{0.48\textwidth}{%
\centerline{(a)}%
}%
\hfill%
\parbox[t]{0.48\textwidth}{%
\centerline{(b)}%
}%
\caption{The real part of exponent $\omega_1$ in the frustrated FP for  $N=2$ at  $\alpha=5$ (a)
and  $N=3$ at  $\alpha=6$ (b) as a function of  $b$   at four
(grey dashed curve (green online)), five (dark dashed curves (blue
online)) and six (solid curves) loops.}
\label{cub_sp}
\end{figure}

We now study the FP {\bf S}${\rm _{cub}}$.
First, we  note that {\bf S}${\rm _{cub}}$  for $N=2$ and $N=3$  is an attractive  focus, that is $\omega_1$ and $\omega_2$ are complex conjugate and  Re($\omega_1$)=Re($\omega_2)>0$.  Our results are shown in  figure~\ref{cub_sp}.
The real part of $\omega_i$ does not show any stationary values as
function of resummation parameters (see figure~\ref{cub_sp}),
 and its behavior is similar to what was obtained for $\omega$ at {\bf S} in the $O(N)$ model. Moreover, the difference of
values obtained in successive loop orders is always more than 0.3, that is 300 times more than for the FP {\bf M}.
It is thus clear that  {\bf S}${\rm _{cub}}$ is a  spurious FP as is the FP {\bf S} for the $O(N)$
model.

\section{Frustrated magnets \label{IV}}

As in the previous section, we consider a model with two couplings but the one that describes
 frustrated antiferromagnets with non-collinear ordering. It is,  for instance, used to describe  a wide class of
magnetic systems, such as antiferromagnets on stacked triangular lattice
or helimagnets (see e.g. \cite{delamotte04,kawamura88}).  Contrarily to
the previously described cases the physics of the phase transition in this
model  is not yet settled. For  $N=2$ and $3$, different approaches predict a first order transition
 \cite{delamotte04}  whereas it is found to be of second order in  \cite{Pelissetto01}.
 The study of this problem within minimal subtraction scheme
  \cite{delamotte06,delamotte08} already shed light on the reason of the discrepancy between these
 approaches showing that the  stable FP should be considered
 spurious. We present here, within the massive RG scheme,
 additional arguments  in this direction.

The effective Hamiltonian  relevant for frustrated
 systems is given by:
\begin{equation}
\begin{array}{ll}
\displaystyle \hspace{0cm}{\mathcal H}{=} \int{\rm d^d} x
\Bigg\{\frac{1}{2} \left[(\partial\phi_1)^2{+}
 (\partial\phi_2)^2                     +                  m^2
 (\phi_1^2{+}\phi_2^2)\right]{+}\frac{u_0}{4!}\ \left[\phi_1^2{+}
\phi_2^2\right]^2 {+}\frac{v_0}{4!}\ \left[(\phi_1
\cdot \phi_2)^2{-} \phi_1^2\,\phi_2^2\right]
\Bigg \}
\end{array}
\label{landau}
\end{equation}
 where the  $\phi_i$, $i=1,2$, are $N$-component vector fields
 and  $u_0$ and $v_0$ are  bare couplings that satisfy $u_0>0$
 and $v_0<4 u_0$ -- which  corresponds to the region where the Hamiltonian is bounded from
 below. For $m^2>0$ the ground state  of Hamiltonian (\ref{landau}) is
 given  by $\phi_1=\phi_2=0$ while  for  $m^2<0$ it is
 given by a configuration  where $\phi_1$ and $\phi_2$
 are orthogonal with   the same norm.   The Hamiltonian (\ref{landau})
 thus describes a symmetry breaking scheme between a disordered and an
 ordered phase where  the $O(N)$ rotation group    is  broken down to
 $O(N-2)$ (for details see
 \cite{delamotte04} for instance).

Let us  first recall  the  picture of FPs obtained  at leading order in $\varepsilon$
 \cite{kawamura88,garel76,bailin77,yosefin85}. The FPs have two
coordinates, $u$ and $v$. For $N$ larger than a critical value $N_{\mathrm{c}}(d)$,
four FPs exist: apart from the usual Gaussian ($u^*=v^*=0$) and
$O(2N)$ ($u^*\ne 0, v^*=0$) FPs, one finds the chiral-antichiral
pair of FPs with coordinates $u^*> 0$ and $v^*> 0$. The chiral FP
{\bf C}$^+$ is stable, whereas the other one, antichiral FP {\bf C}$^-$, is an
unstable one. Above $N_{\mathrm{c}}(d)$, the transition is thus predicted to
be of the second order for systems whose bare couplings lie in the basin of attraction of
{\bf C}$^+$. When  $N$ is lowered at fixed $d$,  {\bf C}$^+$ and {\bf C}$^-$ become
closer and closer and finally collide and disappear at
$N=N_{\mathrm{c}}(d)$. Below $N_{\mathrm{c}}(d)$, there is no longer any stable FP and the
transition is expected to be of  first order.  The value of
$N_{\mathrm{c}}(d)$ for $d=3$ was a subject of intensive studies. In
particular, it has been estimated  within the
$\varepsilon$-expansion  \cite{antonenko95,calabrese03c}, within
pseudo $\varepsilon$-expansion  \cite{Holovatch04} and
within a NPRG approach  \cite{delamotte04}. A six-loop
estimate has found  $N_{\mathrm{c}}=6.23\pm 0.21$  \cite{Holovatch04}
within the  pseudo-$\varepsilon$ expansion
 in good agreement with the value $N_{\mathrm{c}}=6.4\pm 0.4$
 \cite{calabrese03b} derived from the  resummed six-loops $\beta$-functions  computed  in $d=3$.
According to this value of $N_{\mathrm{c}}$, the physical systems that correspond to  $N=2$
or $N=3$ cannot undergo a second order phase transition. However a stable FP has been found for these values of $N$
in the fixed $d=3$ approach  \cite{Pelissetto01}.
Since this FP is not found in the $\varepsilon$-expansion it is natural to wonder whether
it is not an artifact of the fixed $d$ approach as it is the case for the FPs {\bf S} and  {\bf S}$\rm _{cub}$
in the $O(N)$ and cubic model, respectively.
Thus, below we  reconsider the $\beta$-functions of the frustrated
model (\ref{landau}) found in $d=3$ at six loops
 \cite{Pelissetto01}.  Our analysis concludes that the
 stable FP found for frustrated systems at $N=2,\,\,3$  \cite{delamotte06,delamotte08} is spurious.

\begin{figure}[ht]
\begin{center}
\includegraphics[width=0.32\textwidth]{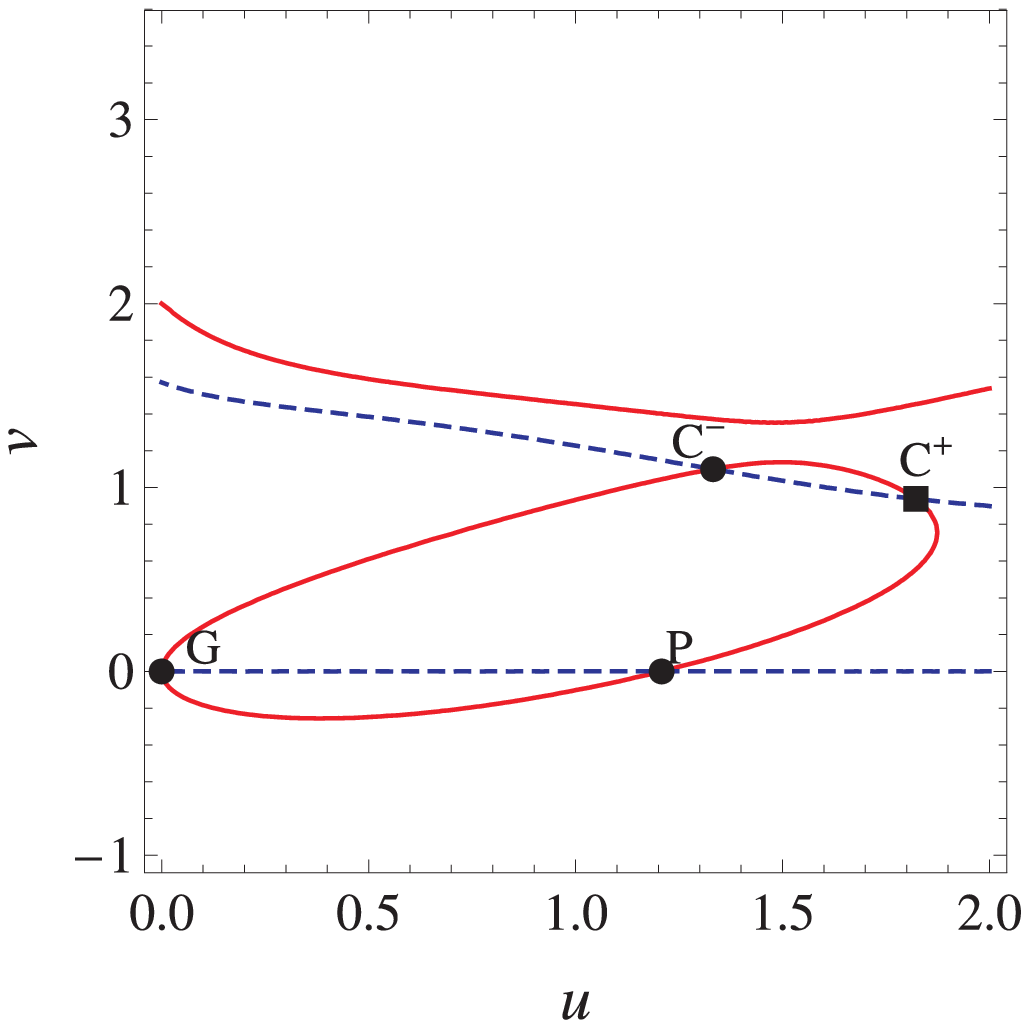}%
\hfill%
\includegraphics[width=0.32\textwidth]{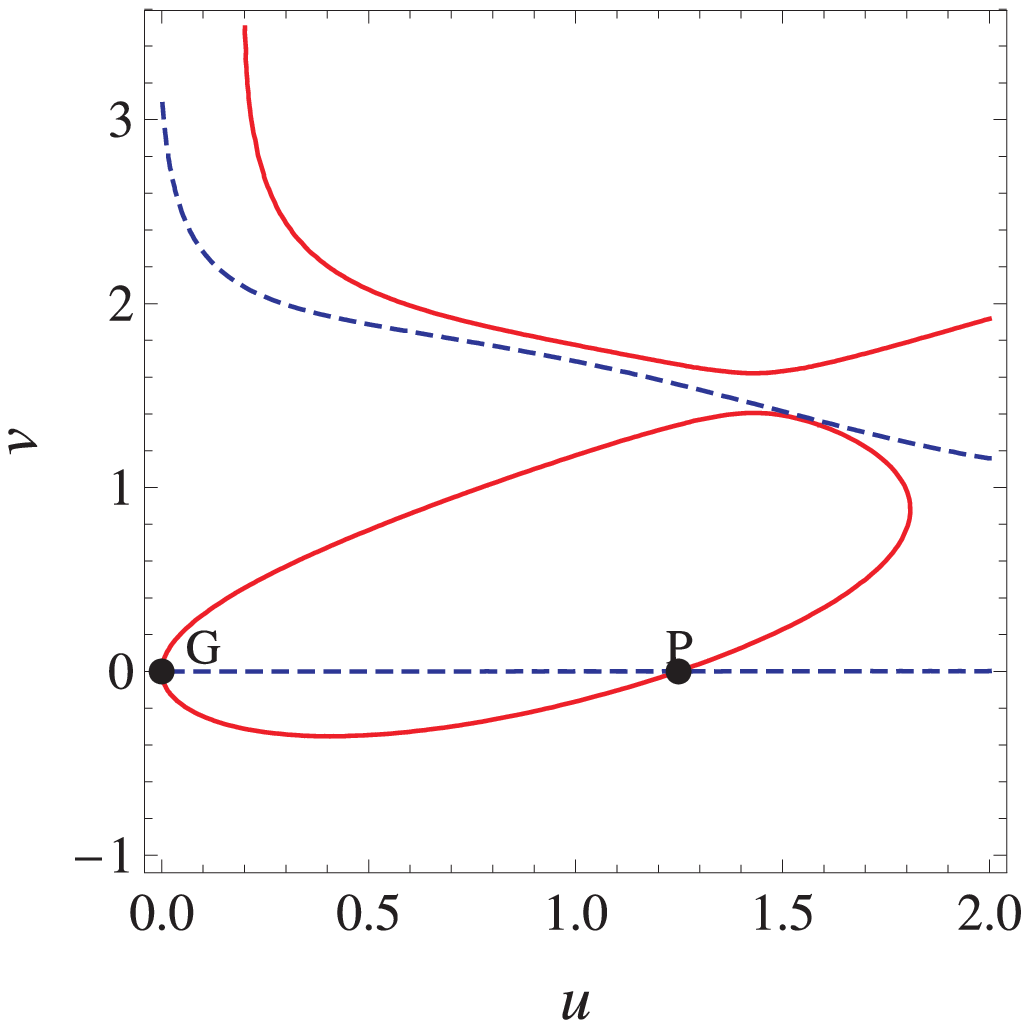}%
\hfill
\includegraphics[width=0.32\textwidth]{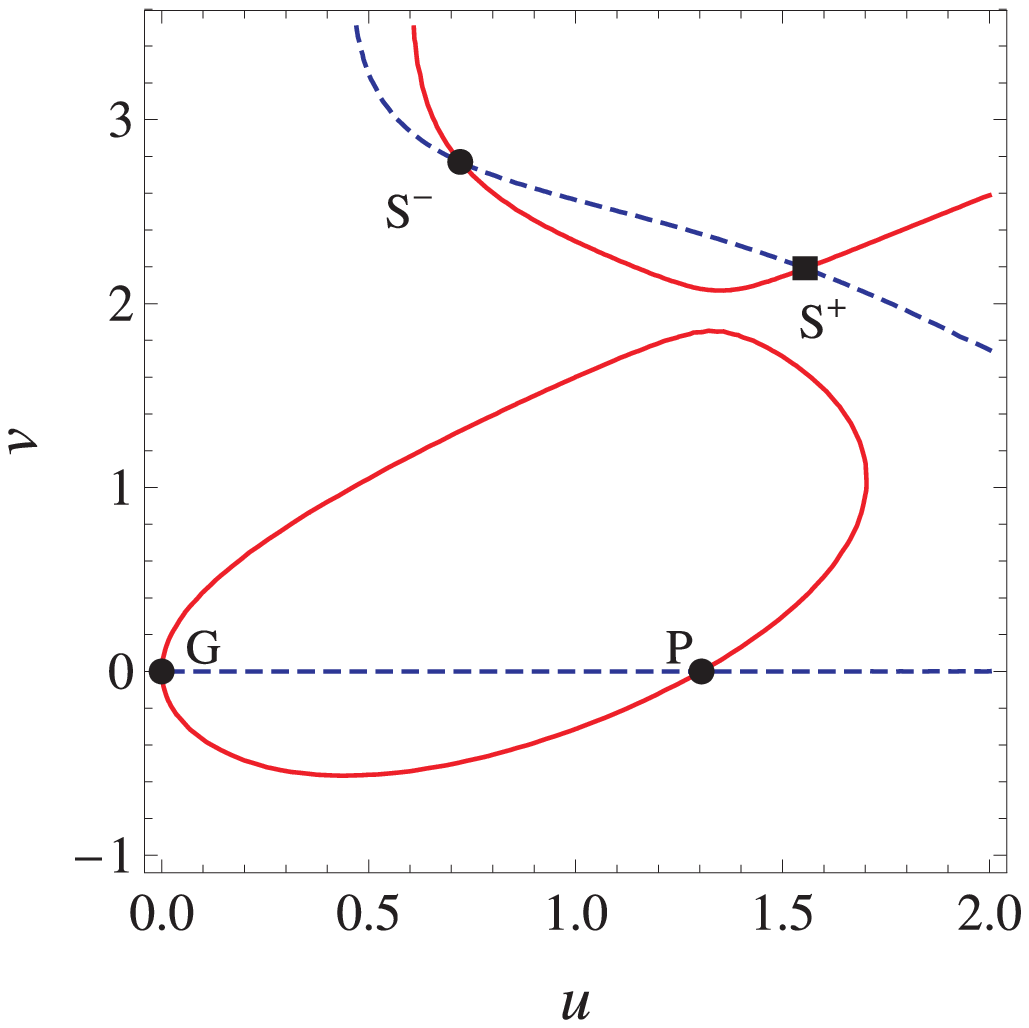}%
\\%
\parbox[t]{0.32\textwidth}{%
\centerline{(a)}%
}%
\hfill%
\parbox[t]{0.32\textwidth}{%
\centerline{(b)}%
}%
\hfill
\parbox[t]{0.32\textwidth}{%
\centerline{(c)}%
}%
\end{center}
\caption{Lines of zeros of $\beta_u$ (solid curves) and
$\beta_v$ (dashed curves) for $N=8$ (a), $N=6$ (b) and $N=4$ (c)  \cite{note1}. Results are obtained
at $\alpha=5$ and $b=15$. The crossings of solid and dashed lines
correspond to different FPs: Gaussian ({\bf G}), Wilson-Fisher
({\bf P}), chiral ({\bf C}$^+$), antichiral ({\bf C}$^-$) and spurious
({\bf S}$^+$, {\bf S}$^-$).} \label{general}
\end{figure}
First we investigate general situation with the change of $N$. Parameter $a$ for frustrated model
has  values depending  on $N$  \cite{Pelissetto01}:
$a=0.14777422\times 9/(2N+8)$ for $4\frac{9}{(2N+8)}>z>0$ and
$a=0.14777422\times( 9/(2N+8)-z/2)$ for $z<0,\quad
z>4\frac{9}{(2N+8)}$. Using this value in calculation  we show the typical behavior
(i.e. the behavior observed for the majority of $\alpha$, $b$) of
the lines of zeros of resummed $\beta$-functions with fixed $d=3$ in figure~\ref{general} for
different $N$, ranging from the large values of $N$, $N>N_{\mathrm{c}}$
(figure~\ref{general}~(a) to the small ones, $N<N_{\mathrm{c}}$
(figure~\ref{general}~(c). As one can see from the figures, the lines
of zeros of  $\beta_u$ form two branches, the upper and the lower
one. Note that the lower branch has a form of a closed contour.
 This contour
corresponds to zeros of function $\beta_u$ which exist already
 in the leading order in $\varepsilon$.  For
$N\geqslant 7$ a curve of zeros of $\beta_v$ intersects the closed contour
generating the chiral-antichiral pair of FPs, {\bf C}$^+$ and {\bf C}$^-$ (see
figure~\ref{general}~(a)).  When values of $N$ are between 7 and 6 this
pair disappears in agreement with the scenario obtained in the
$\varepsilon$-expansion and with an estimate $N_{\mathrm{c}}\approx 6.23$
 \cite{Holovatch04}. Then, till $N\approx 5$ there is no
intersection between curves of zeros $\beta_u$ and $\beta_v$ in
the region $u>0$ and $v>0$ and therefore there are no FPs in the given
region (see figure~\ref{general}~(b)). While at $N<5$ we observe, that
curve of zeros of $\beta_v$ intersects the upper branch of zeros of
$\beta_u$ forming two FPs, stable {\bf S}$^+$ and unstable ones {\bf S}$^-$
(see figure~\ref{general}~(c)). A similar situation has already been observed
in \cite{calabrese03b} that has led  to a conclusion concerning the
existence of another critical value of $N$. From the
above sketched analysis one can see that the FP solutions {\bf C}$^+$, {\bf C}$^-$,
on the one hand and {\bf S}$^+$, {\bf S}$^-$ on the other hand correspond to the
crossing of different branches of the $\beta_u$ curve. In the first
case ({\bf C}$^+$, {\bf C}$^-$) this is the lower branch, whereas in the
second case ({\bf S}$^+$, {\bf S}$^-$) this is the upper one. Below, we will
show the similarities in the behavior of solutions {\bf S}$^+$, {\bf S}$^-$
to those that were obtained in the former section~\ref{III} for
the spurious FP. In the following subsections we separate the analysis
of stable FPs for cases ``large $N$'' and ``small $N$'' meaning by
this two different regimes when $N$ is greater than $N_{\mathrm{c}}$ (c.f.
figure~\ref{general}~(a)) or smaller than $N_{\mathrm{c}}$ (figure~\ref{general}~(c)),
correspondingly.

\subsection{Case of large $N$ }

\begin{figure}[ht]
\hspace{5mm}
\includegraphics[width=0.45\textwidth]{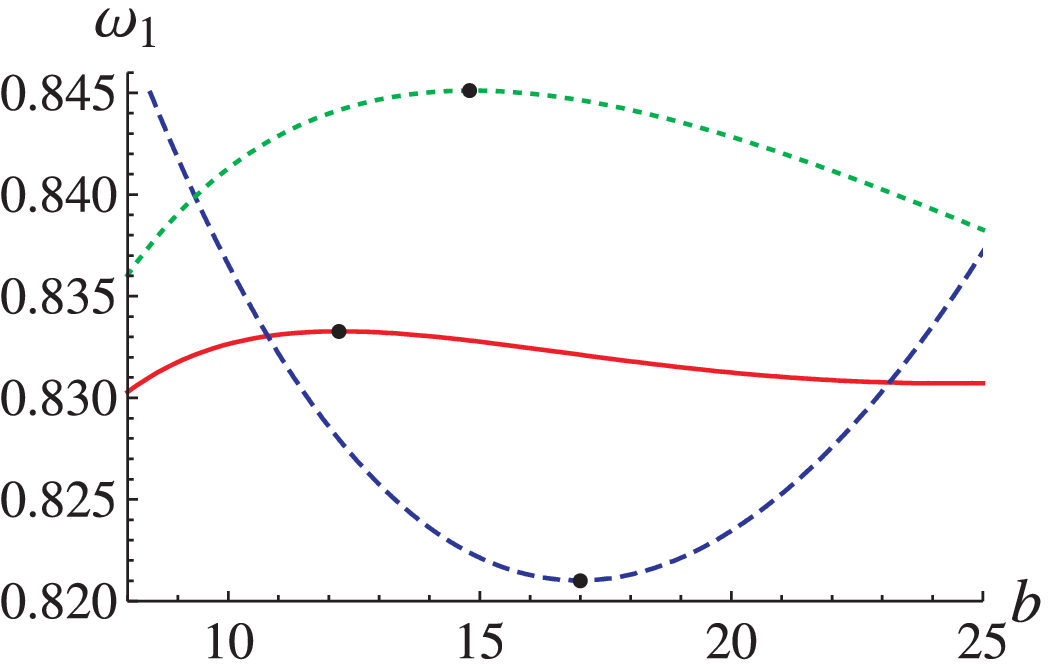}%
\hfill%
\includegraphics[width=0.45\textwidth]{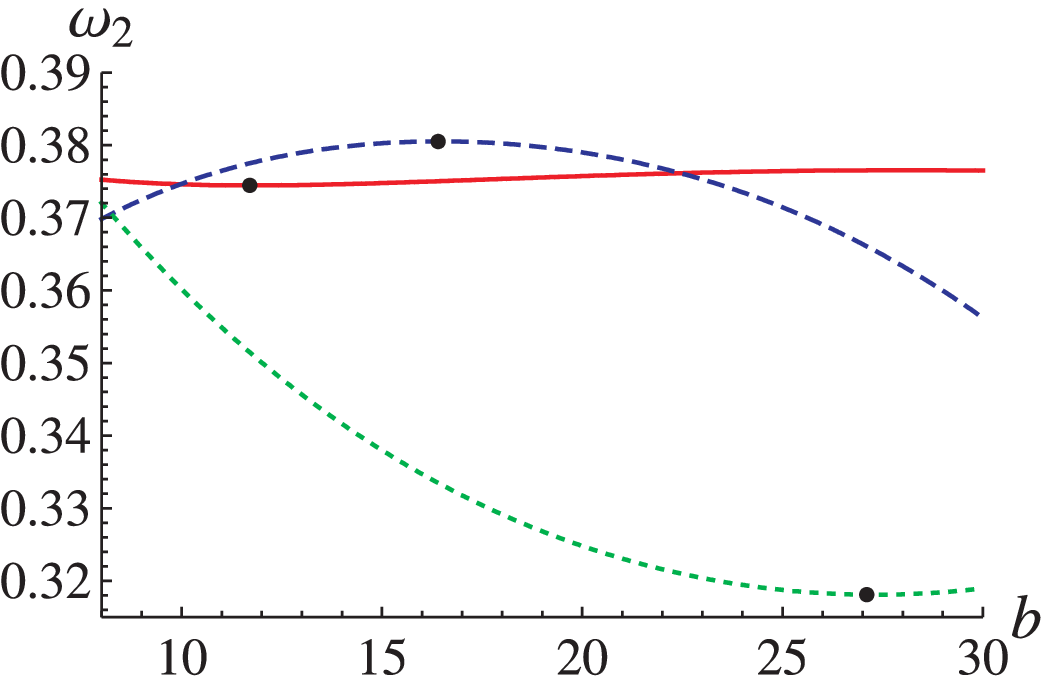}
\hspace{5mm}%
\\%
\parbox[t]{0.48\textwidth}{%
\centerline{(a)}%
}%
\hfill%
\parbox[t]{0.48\textwidth}{%
\centerline{(b)}%
}%
\caption{The exponents $\omega_1$(a) and $\omega_2$ (b) at FP of frustrated model for $N=8$
as a function of  $b$   at  four (grey dashed curves (green online)),
five (dark dashed curves (blue
online)) and six (solid curves) loops. The dots at each loop-order correspond  to a stationary value of
$\omega=\omega(\alpha,b)$ in both  $\alpha$ and $b$ directions at the same time. For $\omega_1$:  $\alpha=5.1$, $b=14.8$ ($L=4$),  $\alpha=7.8$, $b=17$ ($L=5$),  $\alpha=7$, $b=12.2$ ($L=6$); for $\omega_2$:
$\alpha=6.5$, $b=27.1$ ($L=4$),  $\alpha=7.9$,  $b=16.4$ ($L=5$),  $\alpha=7.5$, $b=11.7$ ($L=6$).} \label{frN8}
\end{figure}

Similarly to the above analyzed cubic model here we present the
results of the analysis for stability exponents at $d=3$. We start with the data obtained
 for the stable FP {\bf C}$^+$ at $N>N_{\mathrm{c}}$.
In this case we take $N=8$, therefore $a=0.0554$, and analyze how the
exponent $\omega$ changes with  $\alpha$ and
$b$. We   find that the PMS is satisfied  at  four-, five- and six-loop orders:  for suitable  values of the parameters $b$ and $\alpha$ the two exponents  $\omega_1$ and $\omega_2$  depend weakly on these parameters and are
reasonably well converged. This is clear from figure~\ref{frN8},  where we show the  $b$-dependence of $\omega_1$ and $\omega_2$.  Moreover,  the  difference between the values at five and six loops  of, for instance,  $\omega_2$ is small: $\omega_2(L=6)-\omega_2(L=5)\simeq  0.012$. Note that our values of $\omega_1$ and $\omega_2$  in this  case are fully compatible with those obtained in  \cite{calabrese03b}.

These results indicate that  the convergence properties of  the  $N=8$ frustrated model  are globally similar  to those of the physical FPs of the $O(N)$
 and cubic models although  very likely less accurate  because   in the latter case  the resummation is  less efficient due  to  the presence of complex symmetry.

\subsection{Case of small $N$}

\begin{figure}[ht]
\hspace{5mm}
\includegraphics[width=0.45\textwidth]{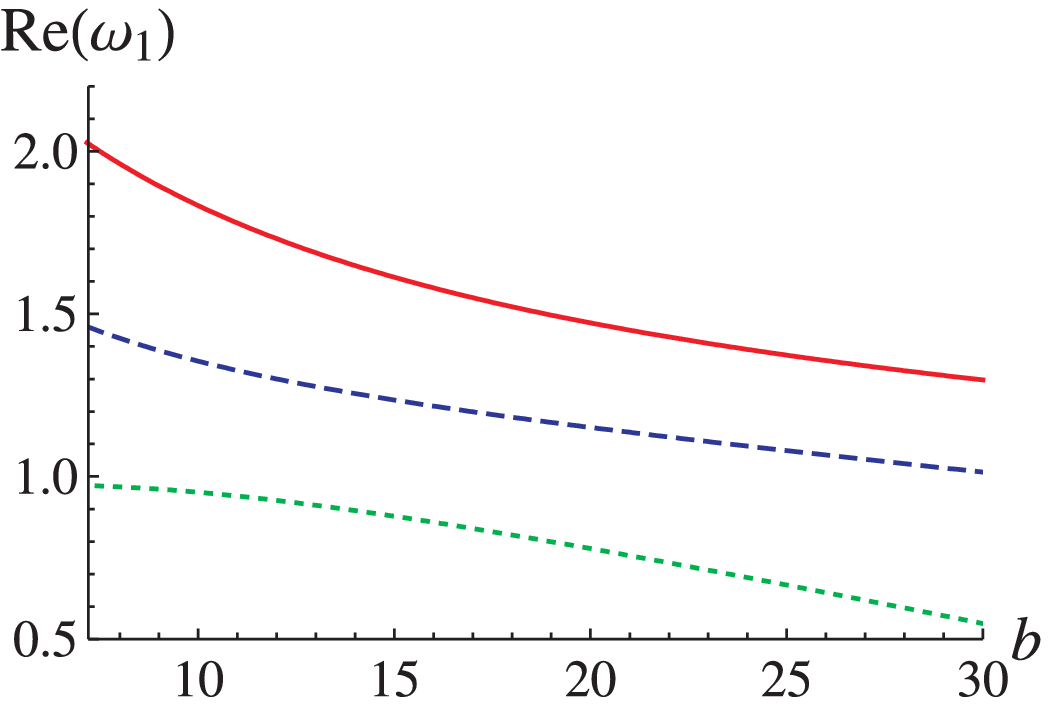}%
\hfill%
\includegraphics[width=0.45\textwidth]{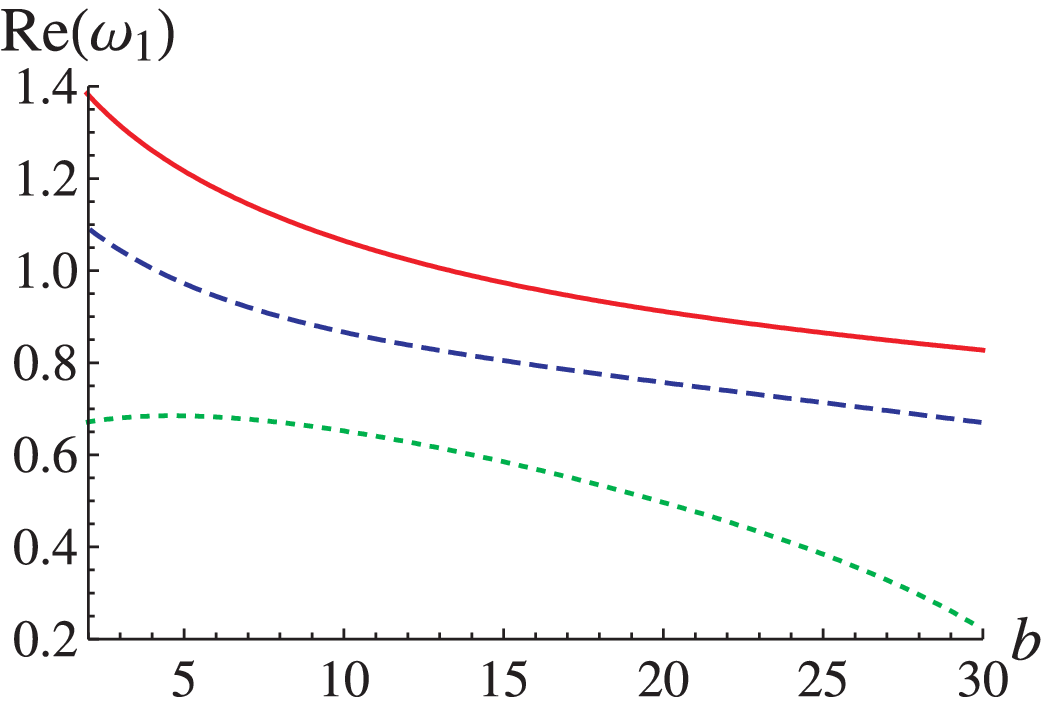}
\hspace{5mm}%
\\%
\parbox[t]{0.48\textwidth}{%
\centerline{(a)}%
}%
\hfill%
\parbox[t]{0.48\textwidth}{%
\centerline{(b)}%
}%
\caption{The real part of the exponent $\omega_1$ in the  FP {\bf S}$^+$ for  $N=2$ (a)
and  $N=3$ (b) as a function of  $b$  for $\alpha=6$  at four
(grey dashed curves (green online)), five (dark dashed curves (blue
online)) and six (solid curves) loops.}
\label{frN2-3b}
\end{figure}

Now, let us study the peculiarities  of the stable FP {\bf S}$^+$ in the
region $u>0,\,v>0$ for the physical cases $N=2,\,3$. The results obtained at the calculation of $\omega$ at this FP with variations for $b$ and $\alpha$ are given in figure~\ref{frN2-3b}.

As it was noted in \cite{Pelissetto01}, stability exponents
attain (in most cases) complex values at FP {\bf S}$^+$ with positive real parts of
$\omega$ indicating   a  stable focus topology. We, again, take for $a$ the value
obtained from the large order analysis  \cite{Pelissetto01}: $a=0.1108$ for $N=2$ and $a=0.095$ for $N=3$.  For these values of $a$ and  for  $L=4,5$ and $6$, we  find that Re($\omega_1$) (or equivalently Re($\omega_2$)) considered as a function of $\alpha$ and
$b$ is    nowhere   stationary, even approximately, see figure~\ref{frN2-3b}. Moreover, at fixed $\alpha$ and
$b$, the gap between the values of Re($\omega_1$) at two successive loop-orders:
Re($\omega_1)(L+1)-$ Re($\omega_1)(L)$, is always large,  of order $0.5$ for $N=2$ and $0.2$ for  $N=3$, see figure~\ref{frN2-3b}. Thus neither the PMS nor the PFAC are  satisfied for these values of $N$.

The $b$-dependence of $\omega_1$ obtained
in figure~\ref{frN2-3b} shows the similarity with the results for
the FP {\bf S} of the $3d$ $O(N)$ model (compare with figure~\ref{omega_O}~(b)) as well as with {\bf S}$\rm _{cub}$  of cubic model (compare with figure~\ref{cub_sp}).  From the figure~\ref{frN2-3b} the bad convergence
properties for spurious FP are evident. Also the
value of $\omega$ in FP {\bf S}$^+$ decreases when  increasing   $b$,
just as it was observed for the FP {\bf S} and {\bf S}$\rm _{cub}$.

Thus we observe a situation analogous to the previously studied models. Similarly to the
 physical  FPs of these models, for FP {\bf C}$^+$ at $N>N_{\mathrm{c}}$,  we observe good convergent
behavior, which is displayed via weak change of exponents and
small difference between the results of different loop order. While
for FP {\bf S}$^+$ at $N<N_{\mathrm{c}}$ we do not observe the regions of stabilization
of critical exponents. The values of stability exponents
considerably differ in different orders. Since such properties are characteristic of FP
{\bf S} and {\bf S}$\rm _{cub}$, this supports the unphysical character of FP {\bf S}$^+$.

\section{Conclusions \label{VI}}
The main motivation of the  study reported in this paper  was to
attract attention to some problems  that arise in the course of RG
analysis of critical behavior of complex Hamiltonians. Being formulated briefly one can
state that RG has proven its accuracy when it is used to calculate
numerical values  of different universal observables that govern the
second order phase transition and critical behavior in more
general sense. However, sometimes RG is used to prove the occurence of
a critical point itself. In such cases, the  occurence of stable and
reachable FP of a RG transformation is considered to be the proof of a
second order phase transition. This is the problem we addressed
in our paper. Namely, given that in principle, a non-physical solution may arise in any
computational scheme, how could
one single-out such solutions from the physical ones?

In particular, several different solutions for FP
may be obtained in  the analysis of the expressions  of
$\beta$-functions obtained within the perturbative RG as series in
coupling constant without use of $\varepsilon$- or
pseudo-$\varepsilon$ expansions  \cite{Parisi73,schloms87} directly
at fixed space dimension $d$. It is clear that the number of such solutions
increases with an increase of the  loop approximation. The way to
distinguish between physical and unphysical FPs is resummation
procedure, which is expected to eliminate all spurious solutions.
However, generally this is not the case. As it follows from the
analysis performed  in this  paper for  the $O(N)$ model, except
 physical Wilson-Fisher FP,   another FP survives.
The current knowledge of the properties of  $O(N)$ model at $d=3$
excludes the occurence of such FP.
Additionally, the obtained FPs clearly differ in their convergent
properties. In particular, the behavior of spurious FP is strongly
dependent on the fit parameters used in resummation and does not satisfy the principles of convergence (PMS, PFAC).
A completely identical picture is observed for the cubic model, whose  critical behavior is also known.
New (spurious) FP, found for this model by a fixed $d$ approach, demonstrates the same behavior as the spurious FP of $O(N)$ model.

Although the problem of discerning between physical and
unphysical FPs is quite general. Here we address it taking as an
example the phase transition for the model of frustrated magnets.
The question of order of phase transition in such systems remains
unclear. The experimental results make it possible
 to divide the conditionally tested materials into two groups that
differ by their critical exponents. The feature of one group is
negative exponent $\eta$, while the scaling relations are violated
for the critical exponents of the other group  (see review in
\cite{delamotte04}). Both phenomena should not occur at the
second order phase transition. Recent numerical simulations give
evidence of the first order phase transition for frustrated systems
 \cite{itakura03,peles04,bekhechi06,zelli07,diep}. The RG investigations based on the
$\varepsilon$- and pseudo-$\varepsilon$-expansions
 \cite{kawamura88,antonenko95,Holovatch04} indicate the absence of
the second order  phase transition within $\phi^4$ theory for a
frustrated model at physical values $N=2$ and $N=3$. This was
corroborated within NPRG  \cite{tissier00,tissier00b,tissier01,delamotte04}. However,
 in the RG analysis at fixed $d$ with application of resummation
 procedures a  stable FP was found at $N=2,\,\,3$, but only starting from four-loop
 approximation  \cite{Pelissetto01}. Since the topology of this FP is
 stable focus, the difference in the experimental data was
 explained by an approach of RG flows to this focus.

The properties of such FP  were already checked within the minimal
 subtraction scheme  \cite{delamotte06,delamotte08}, which makes it possible
 to  continuously follow the behavior of  the FPs with the change of space
 dimension from $d=3$ to $d=4$, where only Gaussian FP describes the
 physics of a system. There, it was shown that the spurious FP
 found at $d=3$ persists at $d=4$ as well. Such an observation
 enabled us to assume that this FP is an unphysical one.

 In the present study the expressions obtained in the massive
 scheme  at $d=3$  are used. Therefore, it is impossible to trace them  to
 $d=4$ by continuous change of $d$. However, the analysis we performed  here
 shows that stable FPs obtained for values $N=2,\,\,3$ on the one hand and for the values higher
 than the critical value $N_{\mathrm{c}}$ one the other hand, differ in their behavior in respect to  a variation
 of fit parameters of the resummation procedure. While for values of
 $N>N_{\mathrm{c}}$ reliable results may be chosen by principles of  convergence, for $N<N_{\mathrm{c}}$ a strong dependence
 of FPs behavior on the resummation parameters does not make it possible to apply these principles. The latter  fact  serves as an argument to consider FPs for $N<N_{\mathrm{c}}$ to be
 unphysical ones.

\section*{Acknowledgements}
We thank Prof. A. Prykarpatsky and Prof. D. Sankovich for an
invitation to submit a paper to this Festschrift. It is our special
pleasure to congratulate Prof. Nikolai N. Bogolubov (Jr.) on the
occasion of his jubilee and to wish him many years of fruitful and
satisfying scientific activity.

We wish to acknowledge the CNRS--NAS Franco-Ukrainian bilateral
exchange program. This work was supported in part by the Austrian
Fonds zur F\"orderung der wissenschaftlichen Forschung under Project
No. P19583--N20.

%
%

\ukrainianpart

\title{Аналіз  3d масивних розвинень пертурбативної ренормалізаційної групи: делікатна справа}
\author{Б. Делямот\refaddr{label1}, М. Дудка\refaddr{label2}, Ю. Головач\refaddr{label2,label3}, Д. Муана\refaddr{label1} }
\addresses{
\addr{label1} Лабораторія теоретичної фізики конденсованих
середовищ, CNRS-UMR 7600, Університет П'єра і Марії Кюрі, 75252
Париж C\'edex 05, Франція \addr{label2} Інститут фізики
конденсованих систем НАН України, UA--79011 Львів, Україна
\addr{label3} Інститут теоретичної фізики, Університет Йогана
Кеплера, A--4040 Лінц, Австрія }
%
%
%

\makeukrtitle

\begin{abstract}
\tolerance=3000%
Аналізується ефективність пертурбативного підходу у методі
ренормалізаційної групи при фіксованій вимірності простору $d$ в
теорії критичних явищ. Розглядається три моделі: $O(N)$, кубічна і
антиферомагнітна модель означена на трикутній ґратці. Ми розглядаємо
усі моделі при фіксованому $d=3$ і аналізуємо процедуру
пересумовування, яка використовується для обчислення критичних
показників. Ми спочатку показуємо, що для $O(N)$ моделі
пересумовування не виключає усі нефізичні (фіктивні) нерухомі точки
(НТ). Тоді уважно вивчається залежність фіктивної НТ, а також НТ
Вільсона-Фішера від параметрів пересумовуваня. Критичні показники в
НТ Вільсона-Фішера слабко  залежать від пареметрів пересумовування.
На противагу цьому, показники в фіктивній НТ, а також саме її
існування, суттєво залежать від цих параметрів. Для кубічної моделі
отримано нову стійку НТ і показано, що  і її властивості також
значно залежать від параметрів пересумовування. Вона виявляється
фіктивною, як очікувалось. Щодо фрустрованої моделі, то існують два
випадки в залежності від значення числа спінових компонент. Коли $N$
більше за критичне значення $N_{\mathrm{c}}$, поведінка у  стійкій нерухомій
точці подібна до поведінки у НТ Вільсона-Фішера. На противагу цьому,
для $N<N_{\mathrm{c}}$ результати отримані для стійкої НТ подібні до тих, що
отримані в фіктивних НТ $O(N)$ і кубічної моделі. З цього аналізу
ми робимо висновок, що стійка НТ знайдена для   $N<N_{\mathrm{c}}$ в
фрустрованій моделі, є фіктивною, Оскільки $N_{\mathrm{c}}>3$, ми робимо
висновок, що перехід для XY і гайзенберґівського фрустрованого
магнетика є фазовим переходом першого роду.
\keywords теорія поля, ренормалізаційна група, критичні явища,
теорія збурень, пересумовування
\end{abstract}


\begin{thebibliography}{100}
\bibitem{note1}
Here and below we use the normalization of couplings and $\beta$-functions in which the coefficient
of the one-loop contribution in $\beta$-functions equals -1
\bibitem{sqrt}
We omit the case of one-loop degenerated $\beta$-functions, as in the
diluted Ising model. In such cases  the two loop contibutios
should be taken into account and Wilson-Fisher FP is found in the
powers of $\sqrt\varepsilon$. See e.g. \protect \cite{Folk03}
\bibitem{note_cub} Value of $\omega_2$ is reported in the erratum  \cite{folk00b} as smallest stability exponent,
while value for largest $\omega_1$ is in the main reference   \cite{folk00b}.
\end{thebibliography}
\end{document}